\documentclass[lettersize,journal]{IEEEtran}
\usepackage{amsmath,amsfonts}
\usepackage{array}
\usepackage[caption=false,font=normalsize,labelfont=sf,textfont=sf]{subfig}
\usepackage{textcomp}
\usepackage{stfloats}
\usepackage{url}
\usepackage{verbatim}
\usepackage{graphicx}
\usepackage{cite}

\usepackage{bm}
\usepackage{multirow}%
\usepackage[ruled]{algorithm2e}%
\usepackage{hyperref}
\usepackage{color}
\usepackage{booktabs}
\usepackage{tabularx}
\usepackage{ragged2e}

\begin{document}

\title{Steering Language Model to Stable Speech Emotion Recognition via Contextual Perception and Chain of Thought}

\author{Zhixian Zhao, Xinfa Zhu, Xinsheng Wang, Shuiyuan Wang, Xuelong Geng, Wenjie Tian, Lei Xie,~\IEEEmembership{Senior Member,~IEEE}

\thanks{Zhixian Zhao, Xinfa Zhu, Shuiyuan Wang, Xuelong Geng, Wenjie Tian, and Lei Xie are with Audio, Speech and Language Processing Group (ASLP@NPU), School of Computer Science, Northwestern Polytechnical University, Xi’an 710072, China (e-mail: zxzhao@mail.nwpu.edu.cn; xfzhu@mail.nwpu.edu.cn; wangshuiyuan@mail.nwpu.edu.cn; xlgeng@mail.nwpu.edu.cn; twj@mail.nwpu.edu.cn; lxie@nwpu.edu.cn).}
\thanks{Xinsheng Wang is with the Hong Kong University of Science and Technology, Hong Kong 999077, China (e-mail: w.xinshawn@gmail.com)}}



\maketitle

\begin{abstract}
Large-scale audio language models (ALMs), such as Qwen2-Audio, are capable of comprehending diverse audio signal, performing audio analysis and generating textual responses. 
However, in speech emotion recognition (SER), ALMs often suffer from hallucinations,  resulting in misclassifications or irrelevant outputs. To address these challenges, we propose $\mathbf{C^2SER}$, a novel ALM designed to enhance the stability and accuracy of SER through \textbf{C}ontextual perception and \textbf{C}hain of Thought (CoT). C$^2$SER integrates the Whisper encoder for semantic perception and Emotion2Vec-S for acoustic perception, where Emotion2Vec-S extends Emotion2Vec with semi-supervised learning to enhance emotional discrimination. Additionally, C$^2$SER employs a CoT approach, processing SER in a step-by-step manner while leveraging speech content and speaking styles to improve recognition. To further enhance stability, C$^2$SER introduces self-distillation from explicit CoT to implicit CoT, mitigating error accumulation and boosting recognition accuracy. Extensive experiments show that C$^2$SER outperforms existing popular ALMs, such as Qwen2-Audio and SECap, delivering more stable and precise emotion recognition. We release the training code, checkpoints, and test sets to facilitate further research\footnote{\href{https://huggingface.co/collections/ASLP-lab/c2ser-67bc735d820403e7969fe8a0}{Hugging Face Collection}, \href{https://github.com/zxzhao0/C2SER}{GitHub Repository}}.
\end{abstract}


\section{Introduction}

\begin{figure}[htb]
  \centering
  \includegraphics[width=0.98\linewidth]{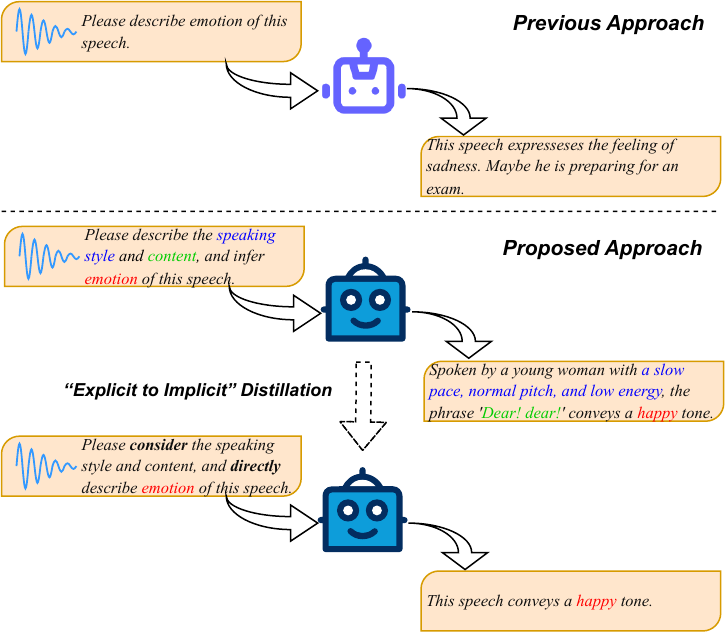}
  \caption {Overview of C$^2$SER. The top path shows how standard models can ``hallucinate" by generating irrelevant context. The bottom path demonstrates our two-step Chain-of-Thought (CoT) approach: first generating a detailed rationale (Explicit CoT), and then internalizing this capability for a direct and stable prediction (Implicit CoT).}
  \label{fig-1}
\end{figure}

\IEEEPARstart{A}{udio} is a multifaceted medium for communication, conveying speech prosody, vocal tone, and paralinguistic cues through its acoustic features. Large-scale audio-language models (ALMs)~\cite{alm-ref,AudioBench,ALM_1} have demonstrated substantial progress in understanding diverse forms of audio signals, which is crucial for advancing Artificial General Intelligence (AGI). With increasing data availability, computational power, and model size, significant strides have been made in speech signal comprehension, analysis, and reasoning, leading to more natural and human-like text responses.

Speech emotions are crucial in communication, influencing how individuals interact and respond through variations in tone, rhythm, and intensity. However, ALMs still face challenges in speech emotion recognition (SER)~\cite{alm2ser,alm2ser2}, often exhibiting ``hallucinations'' that undermine their reliability. These hallucinations involve the generation of any information that is factually ungrounded in the provided audio signal. This issue ranges from the model fabricating plausible but unverified contextual details---such as speculating that a happy speaker is a ``student celebrating university acceptance'' when no such evidence exists---to more critical failures where it fundamentally misinterprets the emotion and invents a baseless justification. As illustrated in Figure~\ref{fig-1}, such a failure can lead to a cheerful utterance being labeled as ``sadness'' with the unsubstantiated rationale that ``Maybe he is preparing for an exam.'' Our work is specifically focused on steering the model's reasoning to be strictly grounded in the acoustic and semantic evidence present in the audio, thereby mitigating these ungrounded outputs.

In this work, we address the hallucination problem in ALM-based SER by incorporating detailed speech information and expanding the model's reasoning length and depth. As illustrated in Figure~\ref{fig-1}, we introduce C$^2$SER, a reasoning specialist ALM designed to improve both the stability and accuracy of SER. 
C$^2$SER integrates two critical components: contextual perception and a chain of thought (CoT), leveraging both speech content and speaking styles (e.g., speaking rate, pitch, energy) to facilitate emotion recognition. 
The encoder of Whisper~\cite{whisper}, trained for automatic speech recognition (ASR), speech translation, and language identification, is employed for semantic perception. For acoustic perception, we introduce Emotion2Vec-S, a refined extension of Emotion2Vec~\cite{Emotion2Vec}, designed to enhance the extraction of emotion-related information from audio. Emotion2Vec-S incorporates semi-supervised contrastive loss at the category level, improving emotional discrimination by combining self-supervised and semi-supervised learning.

Recognizing that speech emotions are influenced by both speech content and speaking styles, such as aggressive speech characterized by a loud volume potentially indicating anger, C$^2$SER employs a CoT~\cite{deepseekr1} training approach to incentivize reasoning capability. This approach decomposes the SER task into sequential steps: first perceiving speech content and speaking style, followed by emotion inference, with the assistance of prior context. This structured method imitates human thinking and reduces the possibility of hallucinations. To further enhance stability and prevent error propagation, especially in longer thought chains, C$^2$SER introduces self-distillation, transferring knowledge from explicit to implicit CoT. This process helps minimize error accumulation, improving the model's overall performance.

We validate the effectiveness of C$^2$SER through extensive experiments using multiple speech corpora and compare C$^2$SER with state-of-the-art models, including SECap~\cite{SECap} and Qwen2-Audio~\cite{Qwen2-Audio}.
To better simulate real-world context, we introduce a new SER test set, Emo-Emilia. Emo-Emilia is created through an automated labeling approach on the in-the-wild Emilia corpus and manually verified to ensure quality and diversity across various scenarios.
Our experimental results demonstrate that C$^2$SER significantly outperforms existing models on public test sets and Emo-Emilia in terms of weighted accuracy (WA), unweighted accuracy (UA), and Macro F1 score, while notably reducing hallucination-related errors. These findings highlight the potential of C$^2$SER to provide stable and reliable emotion recognition in diverse contexts.

The key contributions of our work are summarized as follows:

\begin{itemize} 
\item We propose C$^2$SER, a novel ALM that integrates contextual perception and chain of thought to mitigate hallucinations in SER. 
\item We introduce Emotion2Vec-S, working as the acoustic perception module in C$^2$SER, which enhances the original Emotion2Vec model by incorporating semi-supervised contrastive loss at the category level, greatly improving emotional discrimination. 
\item We conduct comprehensive experiments and introduce a new emotional speech test set Emo-Emilia. The results demonstrate that C$^2$SER outperforms several state-of-the-art ALMs, achieving a more stable SER. 
\item We release the code, checkpoints, and test set to promote further research in the field of SER. 
\end{itemize}

\section{Related Work}

In this section, we review related work that is closely related to C$^2$SER, dividing the discussion into two parts: audio-language models (ALMs) and speech emotion recognition (SER).

\subsection{Audio Language Model}

Large language models (LLMs) have made significant strides in natural language processing (NLP), showcasing remarkable capabilities across a variety of tasks~\cite{LLMsurvey,related_LLM,related_LLM1}. As audio is a critical medium of communication in human interactions and human-computer engagement, recent research has extended LLMs to integrate the audio modality, leading to the development of audio-language models (ALMs). ALMs tackle tasks such as audio event detection, audio captioning, and speech recognition, serving as a cornerstone for comprehensive audio understanding~\cite{qwenaudio,related_kimi,related_osum}.

With the rapid advancements in both LLMs and the audio domain, ALMs have gained significant attention for their powerful general audio comprehension abilities~\cite{related_gama,related_flamingo2}. A typical ALM architecture consists of three core components: an audio encoder for modality-specific feature extraction, an LLM for text generation, and a projection layer to bridge the gap between the audio and text modalities. In addition to these foundational components, several studies have focused on refining ALM performance through innovative model architectures. For example, SALMONN~\cite{SALMONN} utilizes dual encoders to separately process speech and non-speech audio signals, effectively mitigating potential conflicts between different types of audio input. Other approaches have explored training strategies to enhance ALMs' capabilities, with Qwen2-Audio~\cite{Qwen2-Audio} being a notable example. This model employs a comprehensive training pipeline that includes pretraining, supervised fine-tuning (SFT), and reinforcement learning from human feedback (RLHF).

While significant progress has been made in improving the generalization and intelligence of ALMs, their performance in speech emotion recognition (SER) remains unsatisfactory, primarily due to hallucinations. SER is particularly challenging because speech emotions are inherently complex and context-dependent~\cite{alm_AIR-Bench}, making it difficult for ALMs to interpret emotional states accurately. In many cases, the model may be misled by the content of the speech, leading to incorrect classifications or irrelevant responses. Although recent studies like DeSTA~\cite{DeSTA} and Kang et al.~\cite{Kang_Frozen} have attempted to enhance paralinguistic perception through descriptive alignment or frozen LLM paradigms, they do not explicitly address the reasoning deficits that cause such hallucinations.

\subsection{Speech Emotion Recognition}

Speech emotion is a key form of paralinguistic information that effectively conveys the speaker’s intent. Speech Emotion Recognition (SER) aims to automatically identify a speaker’s emotional state from raw audio, with applications in human–computer interaction, healthcare, and affective computing~\cite{ser_related1,ser_related2,ser_related3}. The typical SER pipeline consists of three stages: speech preprocessing, feature extraction, and emotion classification~\cite{review-mer1}. Early studies relied on manually engineered feature sets, such as MFCC, and simple neural network architectures like CNN and RNN, achieving basic performance on laboratory datasets (e.g., CREMA-D~\cite{CREMA-D}, IEMOCAP~\cite{IEMOCAP}).

To address the challenge of recognizing diverse emotional expressions in real-world environments, recent research has shifted towards self-supervised learning (SSL) models~\cite{ssl_ser,ssl_ser2}, known for their powerful generalization capabilities. SSL models are trained on large-scale unlabeled speech data in an unsupervised manner, allowing them to extract rich, generalizable representations directly from raw speech waveforms.  Popular SSL models have demonstrated significant effectiveness in extracting emotional features, serving as robust encoders for SER tasks.
For example, Naini et al.~\cite{ssl_ser3} investigate four SSL models, WavLM~\cite{wavlm}, wav2vec 2.0~\cite{wav2vec2.0}, HuBERT~\cite{Hubert}, and Data2Vec~\cite{data2vec2.0}, and evaluate their cross-domain generalization ability on different speech emotion corpora.
Additionally, researchers have explored emotion-specific SSL models designed to capture emotion-relevant features. A popular approach involves fine-tuning SSL models on emotionally labeled data for specific emotional tasks. A prominent example is Emotion2Vec~\cite{Emotion2Vec}, which is pre-trained on emotional data through self-supervised online distillation. Emotion2Vec uses both utterance-level and frame-level loss as supervision, demonstrating remarkable improvements in emotion recognition across different languages.

Despite the advancements made in SER, mainstream emotional SSL models typically employ single-level constraints, such as utterance-level or category-level constraints. While Emotion2Vec combines utterance-level and frame-level losses that are actually constraints at the utterance level, it still struggles to distinguish similar emotional expressions, such as fear and sadness, potentially leading to confusion in emotion recognition.

\begin{figure*}[htb]
  \centering
  \includegraphics[width=0.78\linewidth]{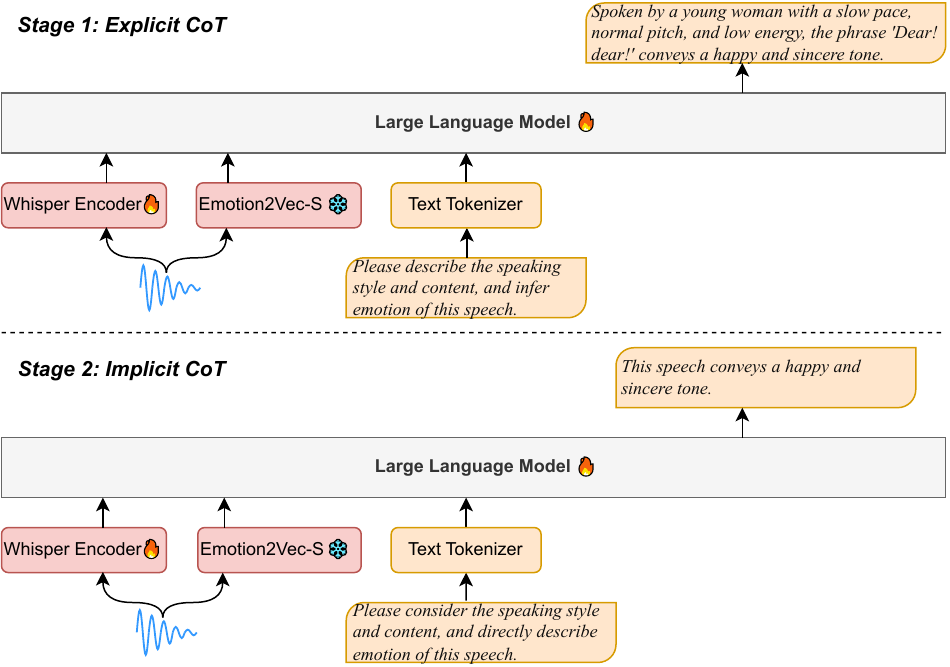}
  \caption {The detailed architecture and two-stage training process of C²SER.
Stage 1 (Explicit CoT): The model is trained to generate a step-by-step rationale by combining semantic features from Whisper and acoustic features from Emotion2Vec-S.
Stage 2 (Implicit CoT): Through self-distillation, the model is trained to produce a direct emotional description, enhancing efficiency while preserving the reasoning capabilities.}
  \label{fig-2}
\end{figure*}

\section{Method}

In this section, we introduce the overall C$^2$SER framework and then detail each of its key components. First, we provide an overview of the architecture and its two main modules. Next, we describe the contextual perception module responsible for extracting semantic and acoustic features. Finally, we present our explicit and implicit chain-of-thought reasoning schemes and explain how they are used to train and refine the model.

\subsection{Framework Overview}
C$^2$SER is designed to mitigate hallucinations in speech emotion recognition (SER) and to deliver stable emotion recognition. As illustrated in Figure~\ref{fig-2}, the C$^2$SER architecture consists of two primary components: a contextual perception module and a text-based large language model (LLM). The contextual perception module extracts detailed information regarding both the semantic and acoustic aspects, which the text LLM subsequently leverages via a chain-of-thought process to make final predictions.

More specifically, the contextual perception module comprises the following elements: a Whisper~\cite{whisper} encoder for semantic perception, Emotion2Vec-S for acoustic perception, and a connection model designed to align the feature dimensions with those required by the text LLM. Formally, given a speech waveform $X$, the Whisper encoder extract semantic representations $S={s_1, s_2, ..., s_N}$ and the Emotion2Vec-S extracts acoustic representations $A={a_1, a_2, ..., a_M}$ from $X$. Let $Y={y_1, y_2, ... , y_T}$ be the text descriptions and $P={p_1, p_2, ... , p_L}$ be the text prompts. The text LLM, parameterized by $\theta$, takes $S$ and $A$ as input and predicts $Y$ in an autoregressive manner. The overall process can be formulated as a conditional probability:
\begin{equation}
    \begin{aligned}
        P(Y | S, A, P; \theta) & = \prod_{t=1}^T P(y_t | s_1,\ldots,s_N,a_1,\ldots,\\
        & a_M, p_1, \ldots, p_L, y_1, \ldots, y_{t-1}; \theta).
    \end{aligned}
\end{equation}

\subsection{Contextual Perception}

Our contextual perception module is designed to extract both semantic and acoustic representations from speech, and it comprises a Whisper encoder and Emotion2Vec-S. Specifically, C$^2$SER employs the Whisper-Medium model as its speech encoder. This model features two one-dimensional convolutional layers with a 2× downsampling factor, followed by 24 Transformer layers. Since Whisper is a supervised model trained for speech recognition and translation, its encoded representations $S$ capture rich semantic information.

Emotion2Vec-S is built upon the universal speech emotion representation model, Emotion2Vec, which follows the architecture of data2vec \cite{data2vec}. Emotion2Vec is pre-trained on open-source, unlabeled emotion data using self-supervised online distillation and has demonstrated superior performance compared to previous state-of-the-art models. It combines two main objectives to learn representations: an utterance-level loss and a frame-level loss. Specifically, the utterance-level loss uses dedicated tokens to learn a representation of the entire utterance's emotion, while the frame-level loss focuses on predicting masked portions of the sequence, forcing the model to learn localized acoustic dependencies. However, a key limitation of Emotion2Vec is that both of these losses operate at the instance level. They lack an explicit mechanism to enforce that embeddings from different utterances of the same emotion category (e.g., two different `fear' samples) should be closer to each other than to embeddings from a different but acoustically similar category (e.g., a `sadness' sample). This can lead to confusion between similar emotional expressions like fear and sadness. To address this limitation, our proposed Emotion2Vec-S extends Emotion2Vec by introducing a category-level contrastive loss, which explicitly pulls embeddings of the same emotion category together while pushing apart those from different categories.

Based on the above observation, Emotion2Vec-S introduces a coarse-level supervision to Emotion2Vec. Vanilla Emotion2Vec expands Data2vec2.0~\cite{data2vec2.0} with a fixed number of utterance tokens and is trained with $\mathcal{L}_{Utt}$ to learn the global emotion and $\mathcal{L}_{Frm}$ to learn the context emotion. Inspired by CLIP~\cite{clip}, Emotion2Vec-S extends Emotion2Vec with a category-level contrastive loss $\mathcal{L}_{Cate}$. Specifically, let $G$ be the global embedding of Emotion2Vec after average pooling. Emotion2Vec-S applies a contrastive loss on $G$ by treating embeddings from utterances of the same emotion category as positive pairs and those from different categories as negative pairs. The model calculates cosine similarities between these embeddings, maximizing the similarity of positive pairs while minimizing that of negative pairs. The overall loss structure of Emotion2Vec-S is illustrated in Figure~\ref{fig:loss_structure}. The total loss function is formulated as follows:

\begin{figure}[t]
  \centering
  \includegraphics[width=\linewidth]{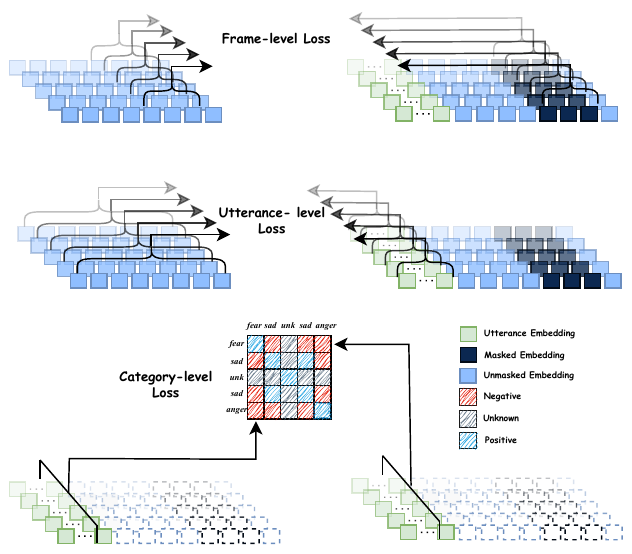}
  \caption {The three types of losses used in Emotion2Vec-s. From top to bottom are utterance-level loss, frame-level loss, and category-level loss.}
  \label{fig:loss_structure}
\end{figure}

\begin{equation}
    \mathcal{L}_{e2v} = \mathcal{L}_{Frm} + \lambda_{utt} \mathcal{L}_{Utt} + \lambda_{cate} \mathcal{L}_{Cate},
\end{equation}
where $\lambda_{utt}$ and $\lambda_{cate}$ are hyperparameters that balance the contributions of the utterance-level and category-level losses, respectively.

\subsection{Explicit Chain-of-Thought}

Explicit CoT reasoning enhances the ability of LLMs to handle specific tasks by detailing intermediate steps, thereby guiding the model through its reasoning process~\cite{excot1,excot2}. In C$^2$SER, explicit CoT is employed to sequentially address the SER task. After the contextual perception module extracts detailed information regarding speech content and speaking styles, C$^2$SER first generates speech transcripts and descriptive captions of speaking styles and then infers the final speech emotion based on the aggregated context.

\textbf{Explicit CoT Data}.The construction of high-quality training data for our explicit CoT framework follows a systematic, three-step process designed to ground the model's reasoning in quantifiable acoustic evidence.

\textit{\textbf{Step 1: Acoustic Attribute Extraction.}} We first employ signal processing tools to extract key acoustic features from each speech waveform. Specifically, we compute pitch contours using the PENN library\textsuperscript{2} and calculate the utterance-level mean to represent the speaker's tone. Energy is extracted using the pyloudnorm library\textsuperscript{3}, yielding a single value for the loudness level. The speaking rate is determined by dividing the number of phonemes in the transcript by the utterance's duration, after trimming silences from the beginning and end of the waveform to ensure accuracy.

\textit{\textbf{Step 2: Feature Discretization.}} To make these continuous acoustic values interpretable for the LLM, we discretize them into categorical labels. After collecting statistics for all utterances in our corpus, we calculate the mean ($\mu$) and standard deviation ($\sigma$) for each attribute. Following the principles of the Central Limit Theorem, we map the values into three levels: `Low' ($\text{value} < \mu - \sigma$), `Medium' ($\mu - \sigma \leq \text{value} \leq \mu + \sigma$), and `High' ($\text{value} > \mu + \sigma$).

\textit{\textbf{Step 3: CoT Path Generation.}} Finally, the discretized acoustic labels (`Low', `Medium', `High'), the original speech transcript, and the ground-truth emotion label are programmatically inserted into a structured prompt. This prompt is then fed to the GLM-4-9B-Chat model, which is instructed to generate a natural-language reasoning path that first describes the speech characteristics and content, and then concludes with an emotion inference. The template for this prompt is shown in Table~\ref{tab:cot-example}.

\begin{table*}[t]
  \centering
  \caption{Template for Explicit CoT data construction, \textcolor{red}{label} will be replaced with the corresponding value.}
  \label{tab:cot-example}
  \begin{tabularx}{\textwidth}{>{\RaggedRight}X|>{\RaggedRight}X}
    \hline
    \textbf{Prompt}: Based on the provided speech features—including a speaking rate of 
      \textless{}\textcolor{red}{speaking rate label}\textgreater{}, 
      a volume level of \textless{}\textcolor{red}{energy label}\textgreater{}, 
      and a pitch of \textless{}\textcolor{red}{pitch label}\textgreater{}—
      along with the text content ‘\textless{}\textcolor{red}{text label}\textgreater{}’
      and the emotion \textcolor{red}{emotion label}, generate a natural and logical emotional description. 
    Here is an example: 
    ‘The speaker spoke at a \textless{}\textcolor{red}{speaking rate label}\textgreater{} pace,
     with a \textless{}\textcolor{red}{pitch label}\textgreater{} tone and 
     \textless{}\textcolor{red}{energy label}\textgreater{} level:
     “\textless{}\textcolor{red}{text label}\textgreater{}”. 
     Based on the analysis of speech characteristics, the emotion was inferred to be 
     \textcolor{red}{emotion label}.’ Ensure including all speech features and logic of the description.
    &
    \textbf{Generated Example}: The speaker spoke at a moderate pace, with a low-pitched tone and a soft volume:
    “together you sort of get this whole narrative of feedback …” Based on the speech characteristics,
    the emotion was inferred to be disgust, revealing a sense of resentment or aversion towards the described situation. \\
    \hline
  \end{tabularx}
\end{table*}

\textbf{Explicit CoT training.} In the explicit CoT training stage, the text LLM is integrated with the contextual perception module, and the entire system is fine-tuned using the explicit CoT data. To further improve the reasoning capabilities of the model, structured text prompts are used, as illustrated in Figure~\ref{fig-2} (stage 1), to guide the model through each intermediate reasoning step. As a result, after this stage of training, C$^2$SER is able to recognize speech content and speaking styles, and subsequently infer the final emotion categories based on the complete speech context.

\subsection{Implicit Chain-of-Thought}

Although the explicit CoT approach enables C$^2$SER to address SER step by step using detailed intermediate representations of speech content and speaking styles, it also introduces inefficiencies during inference and increases the risk of error accumulation~\cite{imcot1,imcot2}. To overcome these limitations, we propose a self-distillation strategy that transitions C$^2$SER from explicit CoT to implicit CoT.

\textbf{Implicit CoT Data.} At this stage, we continue to use the same speech dataset as used in the explicit CoT training; however, the processing of the reasoning path is simplified. Rather than generating detailed intermediate descriptions, the GLM-4-9B-Chat model directly produces descriptions only in terms of emotion labels for each speech segment.

\textbf{Implicit CoT Training.} During this phase, we fine-tune the model on a combination of explicit and implicit CoT data. To ensure that the model maintains its ability to infer emotion categories from rich speech context during the self-distillation process, we gradually transition the training data from explicit to implicit CoT data using a linear schedule. Specifically, we employ a batch-level mixing strategy where the probability of sampling an explicit CoT example decays linearly from 1.0 to 0.0 over the course of this training phase. This ensures that by the end of the training, the model is trained exclusively on implicit CoT data, fully internalizing the reasoning process. Furthermore, we employ customized text prompts, illustrated in Figure~\ref{fig-2} (stage 2), to guide the model reasoning process under the implicit framework. This approach enables C$^2$SER to efficiently generate accurate emotion predictions while addressing the inefficiencies and error propagation associated with explicit CoT.

\section{Data Preparation}

\subsection{Training Data}

\begin{table*}[h]
\centering
\footnotesize
\caption{Statistics of the preprocessed speech corpora used to train C$^2$SER.}
\label{tab:dataset}
\begin{tabular}{llllll}
\toprule
Dataset                & Source  & Emotion Labels Used       & Lang & \#Utts & \#Hrs   \\ \midrule
IEMOCAP                & Act     & Anger, Happiness, Neutral, Sadness   & English & 5331   & 7.0     \\
ESD                    & Act     & Anger, Happiness, Neutral, Sadness, Surprise       & Mix     & 3500   & 29.1    \\
MER2024                & TV      & Anger, Happiness, Neutral, Sadness, Surprise   & Chinese & 5030   & 5.9     \\
BIIC-Podcast(V1.01)    & Podcast & Anger, Happiness, Neutral, Sadness, Surprise, Disgust, Fear & Chinese & 70000  & 147.43  \\
MSP-Podcast(V1.11)     & Podcast & Anger, Happiness, Neutral, Sadness, Surprise, Disgust, Fear & English & 149307 & 237.94  \\ \hline
Internal dataset(Ours) & /       & Anger, Happiness, Neutral, Sadness, Surprise, Disgust, Fear   & Chinese & 439300 & 788.35  \\ \hline
Total                  & -       & Anger, Happiness, Neutral, Sadness, Surprise, Disgust, Fear  & - & 672668 & 1215.72 \\ \bottomrule
\end{tabular}
\end{table*}

The statistics of the training corpora are summarized in Table~\ref{tab:dataset}, which covers seven emotions: anger, happiness, neutral, sadness, surprise, disgust and fear. We utilize six open-source corpora that contain both emotion and text labels, including IEMOCAP~\cite{IEMOCAP}, MELD~\cite{meld}, MSP-Podcast~\cite{msp}, BIIC-Podcast~\cite{biic}, ESD~\cite{ESD}, and MER2024~\cite{mer2024}, alongside an internal corpus containing text labels for model training. To obtain emotion labels for the internal corpus, we apply using an efficient automated labeling method using Emotion2Vec\footnote{\url{https://huggingface.co/Emotion2Vec/emotion2vec_plus_large}} for annotating speech emotions and GLM-4-9B-Chat\footnote{\url{https://huggingface.co/THUDM/glm-4-9b-chat}}~\cite{glm} for annotating text emotions. We then take the intersection of the two annotations to ensure consistency and reliability. Additionally, we incorporate an internal speech corpus containing approximately 2400 hours of unlabeled data during the training of Emotion2Vec. 
To ensure the quality and consistency of our training corpus, we applied a unified preprocessing pipeline to all datasets. This process included: (1) Label Filtering, where we kept only samples within our seven predefined emotion categories (e.g., merging `excited' into `happiness' for IEMOCAP); and (2) Duration Filtering, where we removed utterances longer than 20 seconds. Furthermore, for datasets with official splits like MELD, we used only the training set, and for ESD, we held out 3,500 samples as a separate test set.

\begin{figure}[htb]
  \centering
  \includegraphics[width=0.6\linewidth]{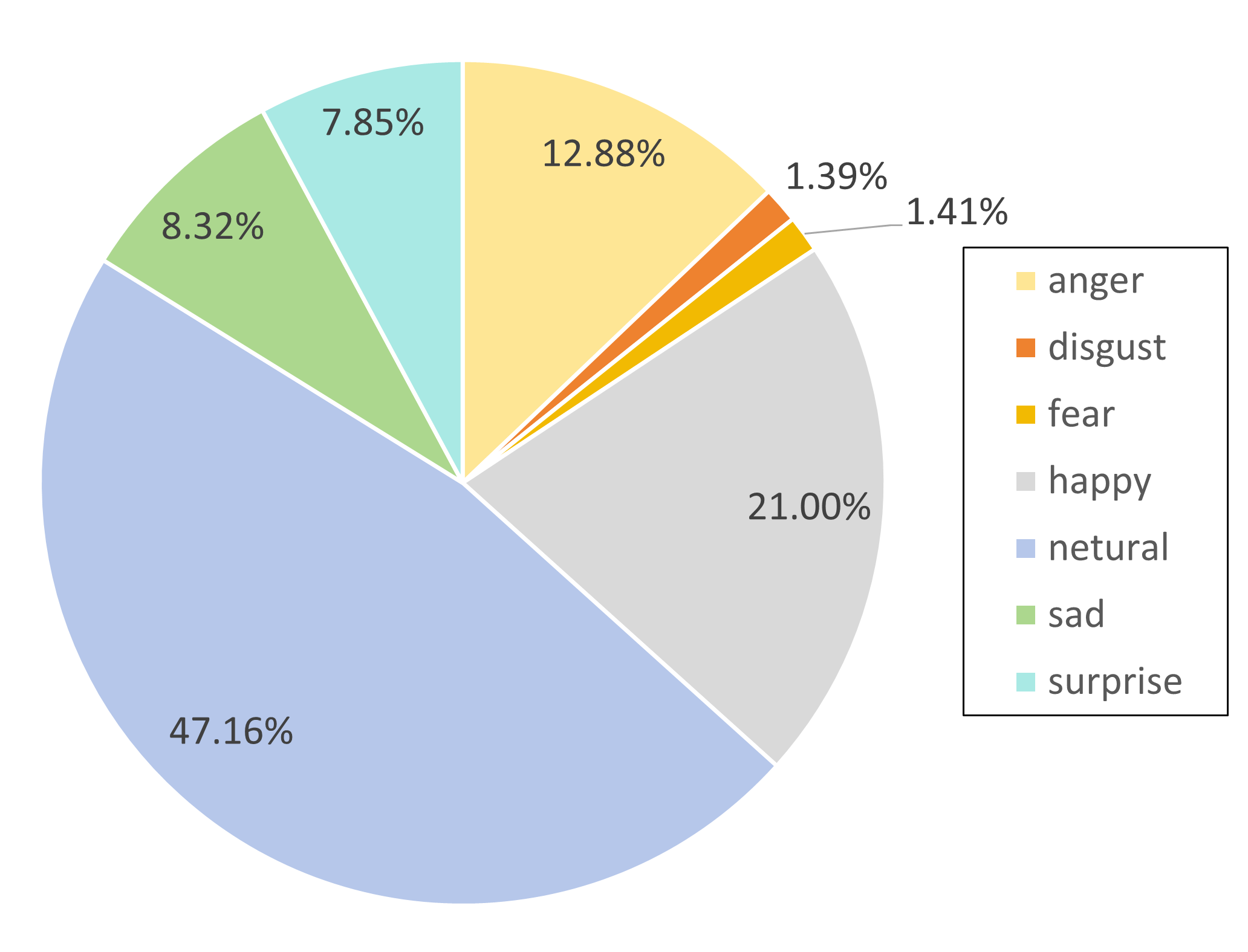}
  \caption{Training data emotion distribution: each slice represents a different emotion, with percentages shown.}
  \label{fig-3}
\end{figure}

\begin{figure}[htb]
  \centering
  \includegraphics[width=0.6\linewidth]{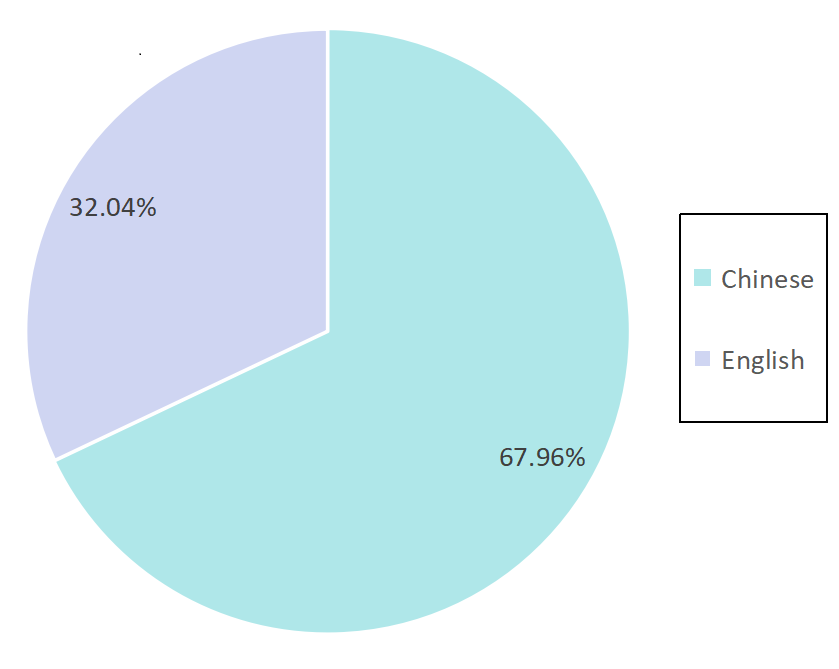}
  \caption{Training data language distribution: each slice represents a different language, with percentages shown.}
  \label{fig-4}
\end{figure}

The distributions of each emotion and the label construction for both explicit and implicit CoT are shown in Figures~\ref{fig-3} and \ref{fig-4}. As observed, neutral emotions account for nearly half of the dataset, while fear and disgust constitute less than 2\%. The scarcity of fear and disgust data arises from challenges such as subjective annotation (e.g., difficulty in accurate identification), limited natural occurrences in contextual expressions, and technical barriers in detecting these emotions from speech or text. Additionally, the proportion of Chinese speech is approximately double that of English speech.

\subsection{Emo-Emilia Dataset}

Considering the complexity of speech emotions in real-world scenarios, we introduce a diverse speech emotion test set, Emo-Emilia. Specifically, we apply the automated labeling approach to annotate Emilia~\cite{emilia}, a large-scale multilingual and diverse speech generation resource with over 100,000 hours of speech data that captures a wide range of emotional contexts. Specifically, we first perform automated annotation on speech and text modalities of the Emilia dataset: employing Emotion2Vec for speech emotion annotation and GLM-4-9B-Chat for text emotion labeling. We subsequently filtered samples with consistent emotion labels across both modalities to ensure annotation accuracy and reliability. From these aligned samples, we randomly selected 300 samples per emotion category covering both Chinese and English languages, resulting in 4,200 candidate data entries. Following this, four bilingual (Chinese-English) speech domain experts independently reviewed all samples, retaining only those with unanimous annotations from all reviewers to enhance data quality and consistency. The final test set, ``Emo-Emilia", comprises 1,400 samples, with 100 samples per emotion category across seven types in both Chinese and English (700 samples per language). The total duration of this test set amounts to approximately 3.3 hours, with 1.4 hours from Chinese speech and 1.9 hours from English speech. Detailed statistics are provided in Table~\ref{tab:emo_emilia_stats}.

\begin{table}[]
\centering
\caption{Statistics of the Emo-Emilia dataset. Each emotion category contains 100 Chinese (CN) and 100 English (EN) samples. Durations are in HH:MM:SS format.}
\label{tab:emo_emilia_stats}
\begin{tabular}{@{}lccc@{}}
\toprule
Emotion   & \# of samples     & CN Duration & EN Duration \\ \midrule
Anger     & 100CN+100EN       & 00:14:19                 & 00:17:23                 \\
Happiness & 100CN+100EN       & 00:12:19                 & 00:17:52                 \\
Neutral   & 100CN+100EN       & 00:13:25                 & 00:11:50                 \\
Sadness   & 100CN+100EN       & 00:13:14                 & 00:20:08                 \\
Surprise  & 100CN+100EN       & 00:12:01                 & 00:13:57                 \\
Disgust   & 100CN+100EN       & 00:10:35                 & 00:15:46                 \\
Fear      & 100CN+100EN       & 00:14:19                 & 00:14:38                 \\ \midrule
Total     & 1400(700CN+700EN) & 01:25:44                 & 01:51:37                 \\ \bottomrule
\end{tabular}
\end{table}

\subsection{Evaluation Benchmarks}

To comprehensively evaluate the model's performance in downstream tasks, we follow the data allocation strategy of EmoBox~\cite{emobox} and use multiple publicly available datasets spanning multiple languages and usage scenarios: Chinese corpora CASIA~\cite{casia} and M3ED~\cite{m3ed}, covering both studio-recorded and spontaneous speech; English corpora MELD~\cite{meld} and EmoV-DB~\cite{emov-db}, representing conversational and acted settings; multilingual corpora ESD~\cite{ESD}, ASVP-ESD~\cite{asvp-esd}, and our proposed Emo-Emilia test set, which provide parallel Chinese–English annotations; and, for zero-shot cross-lingual evaluation, the Italian EMOVO~\cite{emovo} and Mexican MESD~\cite{mesd} datasets. This selection ensures broad linguistic diversity (monolingual to multilingual) and scenario diversity (acted, conversational, studio, spontaneous), enabling a systematic and robust assessment of model generalization across languages and real-world conditions.
 Detailed statistics for these evaluation datasets, including utterance counts, language distribution, and emotion categories, are summarized in Table~\ref{tab:eval_datasets}.

\begin{table*}[h]
\centering
\footnotesize
\caption{Details of the evaluation datasets.}
\label{tab:eval_datasets}
\begin{tabular}{lllll}
\toprule
Dataset    & Source & Emotion Labels                                                & Lang     & \#Utterances \\ \midrule
CASIA      & Act    & Anger, Happiness, Neutral, Sadness, Surprise, Fear            & Mandarin & 1200         \\
M3ED       & TV     & Anger, Happiness, Neutral, Sadness, Surprise, Disgust, Fear   & Mandarin & 24437        \\
MELD       & TV     & Anger, Happiness, Neutral, Sadness, Surprise, Disgust, Fear   & English  & 13706        \\
EmoV-DB    & Act    & Anger, Happiness, Neutral, Disgust, Sleepy                    & English  & 6887         \\
ESD        & Act    & Anger, Happiness, Neutral, Sadness, Surprise                  & Mix      & 35000        \\
ASVP-ESD   & Media  & Anger, Happiness, Neutral, Sadness, Surprise                  & Mix      & 13964        \\
EMOVO      & Act    & Anger, Happiness, Neutral, Sadness, Surprise, Disgust, Fear   & Italian  & 588          \\
MESD       & Act    & Anger, Happiness, Neutral, Sadness, Disgust, Fear             & Mexican  & 862          \\
Emo-Emilia & Media  & Anger, Happiness, Neutral, Sadness, Surprise, Disgust, Fear   & Mix      & 1400         \\ \bottomrule
\end{tabular}
\end{table*}
\section{Experiment Setup}

\subsection{Implement Details}

The architecture of Emotion2Vec-S is based on the original Emotion2Vec\footnote{\url{https://huggingface.co/emotion2vec/emotion2vec_base}} model, with the addition of a classifier consisting of three fully connected layers. To ensure a fair comparison and isolate the gains from our proposed training method, we maintain the exact same backbone architecture, model size, and feature dimensions as the Emotion2Vec model. $\lambda_{utt}$ and $\lambda_{cate}$ are set to 0.1 and 100, respectively. The value of $\lambda_{cate}$ was empirically determined to balance the differing numerical scales of the loss components and was found to yield the best trade-off between training stability and discriminative performance. We employ the Whisper-medium\footnote{\url{https://huggingface.co/openai/whisper-medium}} encoder for semantic feature extraction. The connection module is composed of a 4-layer Transformer followed by a linear layer, with intermediate feature dimensions set to 2,560 in the feed-forward module. For the text LLM component, we utilize the Qwen2-7B-Instruct\footnote{\url{https://huggingface.co/Qwen/Qwen2-7B-Instruct}} model~\cite{Qwen2-Audio} and fine-tune it using Low-Rank Adaptation (LoRA)~\cite{lora}. The LoRA rank is set to 8, the scaling factor is 32, and the dropout rate for LoRA matrices is 0.1.

To train Emotion2Vec-S, we employ 8 Nvidia 4090 GPUs, with a gradient accumulation step set to 2. The optimizer used is Adam, with a learning rate of \(7.5 \times 10^{-5}\) and a weight decay of \(1 \times 10^{-2}\). The learning rate scheduler follows a cosine annealing strategy with a warm-up ratio of 5\%. The remaining hyperparameters are consistent with those used in the vanilla Emotion2Vec model.
To train the entire C$^2$SER model, we utilize 2 Nvidia A6000 GPUs and employ the AdamW optimizer with the following parameters: \(\beta_1 = 0.9\) and \(\beta_2 = 0.99\). The initial learning rate is \(5.0 \times 10^{-5}\), with a weight decay coefficient of 0.01. The learning rate scheduler uses WarmupLR, with the warm-up steps set to 15\% of the total training steps. During C$^2$SER training, Emotion2Vec-S is frozen to retain its ability to effectively extract emotional features from speech.

\subsection{Comparison Systems and Evaluation Details}
\subsubsection{Comparison Systems}
We conduct comparative experiments with several advanced models.
For self-supervised learning (SSL) models (e.g., WavLM~\cite{wavlm}, Data2Vec~\cite{data2vec}, Data2Vec 2.0~\cite{data2vec2.0}), we follow the methodology of EmoBox~\cite{emobox}. First, features are extracted from the last Transformer layer of the pre-trained models and undergo uniform layer normalization to accelerate convergence. Then, a downstream network is applied to perform the SER task, which consists of a simple linear hidden layer, a ReLU activation function, a pooling layer, and a classification head. To ensure a fair comparison, we select models from EmoBox with a comparable parameter scale to that of Emotion2Vec-S, thereby controlling for model size in performance evaluation.

For ALMs, to assess the performance of C$^2$SER, we compare with the following systems.

\textbf{Qwen2-Audio}: A multimodal framework for comprehensive audio understanding and generation. Qwen2-Audio employs Whisper-large-V3 as the audio encoder to capture subtle acoustic features and integrates the Qwen-7B LLM as the foundational component, enabling efficient alignment and generation between audio and text.

\textbf{SenseVoice-Small}: An encoder-only speech foundation model designed for rapid voice understanding. It employs a memory-equipped self-attention network (SAN-M) to enable fast and efficient inference.

\textbf{SECap}: A framework that generates high-quality style captions. It uses HuBERT to extract speech features, Q-Former as the Bridge-Net, and LLaMA as the text decoder to produce coherent style captions. We train SECap on the same data as C$^2$SER.

In addition to these end-to-end models, we introduce a strong cascaded system baseline to specifically validate the advantages of our unified framework.

\textbf{Whisper-m + Qwen2-7B-Instruct}: This system first transcribes speech into text using the Whisper-medium model and then feeds the plain text into a standalone Qwen2-7B-Instruct model for emotion recognition. This baseline represents an approach that relies solely on the semantic content of speech, allowing us to quantify the benefits of direct acoustic feature integration.

Our evaluation strategy for these systems varies based on their original design to ensure a fair and rigorous comparison. For foundation models with built-in SER functionality like Qwen2-Audio and SenseVoice-Small, we follow standard practice by directly evaluating their official checkpoints. Conversely, for a non-SER architecture like SECap, we train it on our speech corpora to establish a strong SER baseline.

\subsubsection{Evaluation Details} We evaluate the models using three key metrics: weighted average accuracy (WA), unweighted average accuracy (UA), and the Macro F1 score. WA represents the overall accuracy of the model, UA corresponds to the average class-wise accuracy, and the Macro F1 score provides a balanced evaluation, particularly useful in cases of class imbalance. 

For all test sets, we first harmonize emotion labels into seven core categories (e.g., merging `amused,' `joy,' and `happy' into the happiness category). The evaluation protocols then differ based on the model type: We test SSL models using leave-one-session-out five-fold cross-validation, following the EmoBox protocol. For ALMs, we conduct inference on designated test sets. Specifically, for datasets with official splits like MELD, we evaluate performance on the official test set. For the ESD dataset, we use the held-out set of 3,500 utterances that were not part of the training data. Since the output of ALMs can be free-form text, we then employ Qwen2.5-14B-Chat\footnote{\url{https://huggingface.co/Qwen/Qwen2.5-14B-Instruct}} to directly extract the most appropriate emotion labels from the descriptions generated by ALMs. We use the following prompt:
\textit{“Given the following text, determine its corresponding emotion and output only the single most appropriate emotion label. The possible labels are: anger, happiness, neutral, sadness, surprise, disgust, fear”}

\begin{table*}[]
\centering
\caption{Emotion2Vec-S performance on datasets of Chinese, English, Italian, Mexican and mixed languages. The best and the second best result is shown in \textbf{bold} and by \underline{underlined}.}
\label{tab:ssl_result}
\resizebox{\textwidth}{!}
{
\begin{tabular}{l|ccc|ccc|ccc}
\hline
\multirow{2}{*}{Model} & UA(\%) ↑             & WA(\%) ↑             & F1(\%) ↑              & UA(\%) ↑             & WA(\%) ↑             & F1(\%) ↑              & UA(\%) ↑             & WA(\%) ↑             & F1(\%) ↑             \\ \cline{2-10} 
                       & \multicolumn{3}{c|}{M3ED (Chinese)}                                       & \multicolumn{3}{c|}{MELD (English)}                                       & \multicolumn{3}{c}{ESD (Mixlingual)}                                       \\ \hline
WavLM-base             & 22.76    & 42.79    & 22.03    & 23.44          & 44.71          & 24.25          & 72.90    & 72.90    & 72.55                    \\
data2vec base              & 19.44    & 37.32    & 19.24    & \underline{23.82}          & \underline{45.57}          & \underline{24.37}          & 65.05    & 65.05    & 64.55                    \\
data2vec2.0 base          & \underline{22.82}    & 41.42    & \underline{22.89}    & \textbf{24.79}          & \textbf{46.65}          & \textbf{25.28}          & \underline{73.40}    & \underline{73.40}    & \underline{73.10}                    \\
Emotion2Vec            & 22.04     & \underline{48.28}     & 20.79     & 23.20          & 44.96          & 24.05          & 70.22    & 70.22    & 70.06                    \\
Emotion2Vec-F          & 20.80    & 48.14    & 18.35    & 19.91          & 41.82          & 20.35          & 64.18    & 64.18    & 63.96                    \\
Emotion2Vec-S          & \textbf{23.82}     & \textbf{50.21}     & \textbf{23.13}     & 21.31          & 45.38          & 21.77          & \textbf{79.84}    & \textbf{79.84}    & \textbf{79.72}                    \\ \hline
Model                  & \multicolumn{3}{c|}{CASIA (Chinese)}                                      & \multicolumn{3}{c|}{EmoV-DB (English)}                                    & \multicolumn{3}{c}{ASVP-ESD (Mixlingual)}                                  \\ \hline
WavLM-base             & 47.25    & 47.25    & 41.78    & \textbf{98.38}     & \textbf{98.49}     & \textbf{98.39}    & 46.38     & 58.05     & 47.35     \\
data2vec base              & 34.72    & 34.72    & 30.88    & 93.26     & 93.61     & 93.23    & 37.66     & 50.79     & 38.26     \\
data2vec2.0 base          & 43.31    & 43.31    & 38.90    & 95.81     & 96.09     & 95.80    & 46.0        & 57.57     & 46.62     \\
Emotion2Vec            & \underline{47.58}    & \underline{47.58}    & \underline{43.55}    & 96.71     & 96.90     & 96.74    & \textbf{48.60}      & \underline{58.30}      & \textbf{49.60}     \\
Emotion2Vec-F          & 43.18    & 43.18    & 39.28    & 96.68     & 96.94     & 96.70    & 47.05     & 57.77     & \underline{48.25}     \\
Emotion2Vec-S          & \textbf{62.95}    & \textbf{62.95}    & \textbf{60.2}     & \underline{97.04}     & \underline{97.30}     & \underline{97.08}    & \underline{48.20}      & \textbf{58.88}      & 45.63      \\ \hline
Model                  & \multicolumn{3}{c|}{EMOVO (Italian)}                                      & \multicolumn{3}{c|}{MESD (Mexican)}                                       & \multicolumn{3}{c}{Emo-Emilia (Mixlingual)}                                     \\ \hline
WavLM-base             & 42.39    & 42.39    & 37.33    & 42.58     & 43.52     & 42.94    & 67.26          & 67.26          & 67.28          \\
data2vec base          & 32.47    & 32.47    & 29.22    & 34.37     & 34.35     & 33.24    & 63.80          & 63.80          & 63.72          \\
data2vec2.0 base      & \textbf{42.96}    & \textbf{42.96}    & \textbf{41.01}    & 44.86     & 44.85     & 43.60     & 64.60          &64.60           & 64.46          \\
Emotion2Vec            & 41.02    & 41.02    & 38.60    & 50.56     & 50.48     & 50.10    & \underline{68.02}          & \underline{68.02}          & \underline{68.00}          \\
Emotion2Vec-F          & 38.65    & 38.65    & 34.83    & \underline{55.38}     & \underline{55.36}     & \underline{55.18}    & 59.24          & 59.24         & 59.00          \\
Emotion2Vec-S          & \underline{42.88}    & \underline{42.88}    & \underline{40.87}    & \textbf{59.57}     & \textbf{59.62}     & \textbf{59.28}    & \textbf{80.66}          & \textbf{80.66}          & \textbf{80.58}          \\ \hline
\end{tabular}}
\end{table*}

\section{Experimental Results}

In this section, we first assess the quality of the Emotion2Vec-S representations by comparing them with other SSL pre-trained models. Next, we evaluate C$^2$SER against leading audio-language models. We then examine category-level accuracies for both Emotion2Vec-S and C$^2$SER, followed by an analysis of the impact of our chain-of-thought training. Finally, an ablation study dissects the contribution of each module within C$^2$SER.

\subsection{Evaluation of Emotion2Vec-S}

The results are presented in Table~\ref{tab:ssl_result}, where we compare Emotion2Vec-S with various SSL pre-trained models of similar model size and training corpora. Notably, Emotion2Vec-F refers to the Emotion2Vec model trained directly on the same corpora as Emotion2Vec-S, allowing us to investigate the impact of different datasets on model performance. The results demonstrate that Emotion2Vec-S consistently outperforms other models across most datasets. Interestingly, the performance gap between Emotion2Vec and Emotion2Vec-S indicates that the training corpus does influence the results, but it does not always lead to significant improvements. Nonetheless, Emotion2Vec-S consistently shows steady improvement compared to both Emotion2Vec-F and Emotion2Vec, validating the effectiveness of semi-supervised contrastive learning.

Furthermore, we observe that the models exhibit varying performance across different test sets and languages. Specifically, Emotion2Vec-S outperforms comparison models by a significant margin on Chinese test sets while achieving competitive results on English datasets. Additionally, Emotion2Vec-S excels in multilingual test sets, particularly in ESD and Emo-Emilia. When extended to other languages, Emotion2Vec-S achieves the best results on the Mexican test set and ranks second on the Italian test set. Although Emotion2Vec-S is trained primarily on Chinese and English speech corpora, these results highlight its impressive generalization capabilities across different languages. Overall, these findings suggest that Emotion2Vec-S offers superior emotion discrimination compared to the original Emotion2Vec model, establishing it as a robust foundation model for extracting speech emotion representations.

\begin{table*}[]
\centering
\caption{C$^2$SER performance on datasets of Chinese, English, Italian, Mexican and mixed languages. The best and the second best result is shown in \textbf{bold} and by \underline{underlined}.}
\label{tab:llm_result}
\resizebox{\textwidth}{!}
{
\begin{tabular}{l|ccc|ccc|ccc}
\hline
\multirow{2}{*}{Model} & UA(\%) ↑             & WA(\%) ↑             & F1(\%) ↑              & UA(\%) ↑             & WA(\%) ↑             & F1(\%) ↑              & UA(\%) ↑             & WA(\%) ↑             & F1(\%) ↑             \\ \cline{2-10} 
                       & \multicolumn{3}{c|}{M3ED (Chinese)}                                       & \multicolumn{3}{c|}{MELD (English)$^{*}$}                                       & \multicolumn{3}{c}{ESD (Mixlingual)$^{*}$}                                       \\ \hline
Qwen2-Audio            & 19.53                & 41.38                & 15.50                 & \underline{36.82}                & \textbf{54.56}                & \textbf{37.33}                 & 56.26                & 56.26                & 33.06                \\
SenseVoice-S           & 23.13    & 23.09    & 21.11    & 0.99      & 1.95      & 1.68     & 52.23                & 52.23                & 42.20                \\
SECap                  & 23.74    & 29.90    & 18.31    & 19.41          & 22.53          & 13.91         & 42.51     & 42.51     & 25.55     \\
Whisper-m + Qwen2-7B                & 23.13 & 45.21 & 23.72 & 36.26 & 52.99 & \underline{36.81} & 9.22 & 64.57 & 11.21 \\
C$^2$SER(Explicit CoT) & \underline{32.29}    & \underline{47.59}    & \underline{26.99}    & 30.36     & 51.39     & 27.45    & \underline{93.81}     & \underline{93.86}     & \underline{68.19}     \\
C$^2$SER(Implicit CoT) & \textbf{36.68 }   & \textbf{50.57}    & \textbf{29.40}    & \textbf{38.66}     & \underline{53.10}     & 33.10    & \textbf{96.33}     & \textbf{96.34}     & \textbf{81.62}     \\ \hline
Model                  & \multicolumn{3}{c|}{CASIA (Chinese)}                                      & \multicolumn{3}{c|}{EmoV-DB (English)$^{*}$}                                    & \multicolumn{3}{c}{ASVP-ESD (Mixlingual)}                                  \\ \hline
Qwen2-Audio            & \underline{48.17}                & \underline{48.17}                & 35.58                 & \textbf{99.28}                & \textbf{99.38}                & \textbf{79.51}                 & \underline{43.44}                & \underline{48.01}                & \textbf{36.53}                \\
SenseVoice-S           & 33.58    & 33.58    & 24.32    & 39.01     & 42.13     & 33.11    & 16.55                & 16.19                & 21.57                \\
SECap                  & 33.75    & 33.75    & 23.54    & 28.85     & 30.26     & 17.10    & 25.07     & 27.95     & 19.42     \\
Whisper-m+Qwen2-7B                & 13.93 & 16.25 & 7.06 & 12.13 & 34.09 & 12.09 & 30.49 & 39.07 & 30.13 \\
C$^2$SER(Explicit CoT) & 46.62    & 46.62    & \underline{37.51}    & 59.07     & 63.18     & 36.55    & 41.62     & 47.34     & 32.58     \\
C$^2$SER(Implicit CoT) & \textbf{53.33}    & \textbf{53.33}    & \textbf{42.85}    & \underline{59.66}     & \underline{63.38}     & \underline{41.63}    & \textbf{43.86}     & \textbf{48.54}     & \underline{34.06}     \\ \hline
Model                  & \multicolumn{3}{c|}{EMOVO (Italian)}                                      & \multicolumn{3}{c|}{MESD (Mexican)}                                       & \multicolumn{3}{c}{Emo-Emilia (Mixlingual)}                                     \\ \hline
Qwen2-Audio            & 35.88                & 35.88                & 26.22                 & 23.60                & 23.55                & 21.62                 & 39.07   & 39.07     & 31.91     \\
SenseVoice-S           & 14.12    & 14.12    & 14.42    & 23.13     & 23.09     & 21.11    & 63.31                 & 63.31                & 56.84                   \\
SECap                  & 26.36    & 26.36    & 17.31    & \underline{28.40}     & \underline{28.39}     & 21.24    & 32.50     & 32.50     & 23.62     \\
Whisper-m+Qwen2-7B                & 15.31 & 15.31 & 7.07 & 22.44 & 26.10 & 19.07 & 63.31 & 67.36 & 60.89 \\
C$^2$SER(Explicit CoT) & \underline{37.59}    & \underline{37.59}    & \underline{27.33}    & 28.15     & 28.09     & \textbf{21.75}    & \underline{68.29}     & \underline{68.29}     & \underline{61.28}     \\
C$^2$SER(Implicit CoT) & \textbf{41.67}    & \textbf{41.67}    & \textbf{35.93}    & \textbf{28.60}     & \textbf{28.54}     & \underline{21.66}    & \textbf{69.00}     & \textbf{69.00}     & \textbf{61.61}     \\ \hline
\end{tabular}}
\begin{minipage}{\textwidth} 
\vspace{2pt} 
\footnotesize{$^{*}$Qwen2-Audio's results on EmoV-DB may indicate data leakage (i.e., inclusion of the dataset in training). Results for MELD and ESD reflect in-domain evaluation. All other datasets were evaluated in a zero-shot, cross-dataset generalization setting.}
\end{minipage}
\end{table*}

\subsection{Evaluation of C$^2$SER}

We compare C$^2$SER with several leading audio-language models (ALMs) across various test sets, with the results presented in Table~\ref{tab:llm_result}. C$^2$SER consistently demonstrates superior performance over other end-to-end models like SECap, which was trained on the same corpora. This superiority is particularly evident when compared to the `Whisper-m + Qwen2-7B-Instruct' cascaded baseline. The cascaded approach suffers from a severe performance degradation on datasets where acoustic features are dominant (e.g., ESD, CASIA), as it completely discards critical paralinguistic cues during the ASR step. In contrast, by processing acoustic and semantic information simultaneously, our end-to-end C$^2$SER framework shows more robust and accurate emotion recognition across a wider range of scenarios. These results strongly validate that incorporating speech context through our unified, chain-of-thought approach effectively improves SER.

Furthermore, within our C$^2$SER framework, we observe significant improvements when advancing from explicit to implicit CoT. This performance gain highlights the success of our self-distillation strategy, which preserves reasoning capabilities while significantly reducing the potential for error accumulation in longer thought chains.

Beyond comparing C$^2$SER to other ALMs, it is crucial to position it relative to the traditional supervised paradigm. A comparison between Table~\ref{tab:llm_result} and Table~\ref{tab:ssl_result} reveals a noteworthy trend: on certain datasets (e.g., EmoV-DB and MESD), the zero-shot performance of C$^2$SER is lower than that of Emotion2Vec-S evaluated with in-domain fine-tuning. This is not a model deficiency but an expected outcome stemming from their fundamental evaluation paradigms. Specifically, Emotion2Vec-S acts as a ``domain expert'' optimized for a specific dataset through supervised cross-validation. In contrast, C$^2$SER acts as a ``reasoning specialist,'' evaluated under zero-shot conditions on these datasets. Therefore, this comparison clearly illustrates the trade-off between performance and generalization: while a specialized ``expert model'' excels with sufficient in-domain data, the value of C$^2$SER lies in its powerful zero-shot generalization capability, making it critically suited for real-world applications where data is sparse or domains constantly shift.

\begin{table*}[t]
  \centering
  \caption{Effectiveness of CoT training on the Emo-Emilia test set.}
  \label{tab:cot-training}
\begin{tabular}{lllll}
\toprule
CoT Phase                       & Model                  & UA(\%) ↑ & WA(\%) ↑ & F1(\%) ↑ \\ \midrule
\multirow{3}{*}{Inference} & Qwen2-Audio  & 39.07    & 39.07    & 31.91    \\
                           & Qwen2-Audio (Explicit CoT)  & 32.57    & 32.57    & 38.12    \\
                           & Qwen2-Audio (Implicit CoT)  & 25.79    & 25.79    & 33.21    \\ \hline
\multirow{2}{*}{Training}  & C$^2$SER (Explicit CoT) & 68.29    & 68.29    & 61.28    \\
                           & C$^2$SER (Implicit CoT) & 69.00    & 69.00    & 61.61    \\ \bottomrule
\end{tabular}
\end{table*}

\begin{table}[t]
\centering
\caption{Ablation study of C$^2$SER on the Emo-Emilia test set.}
\label{tab:ablation}
{
\begin{tabular}{llll}
\toprule
Model                                 & UA(\%)↑ & WA(\%)↑ & F1(\%)↑ \\ \midrule
C$^2$SER              & \textbf{69.00}        & \textbf{69.00}       & \textbf{61.61}        \\
w/o Whisper encoder                         & 32.07        &  32.07      & 34.56        \\
w/o Emotion2Vec-S           & 57.93        &  57.93      & 51.10        \\
w/o CoT & 43.14        & 43.14       & 36.15        \\
\bottomrule
\end{tabular}}
\end{table}

\begin{table}[t]
\centering
\caption{Performance improvement of C$^2$SER after fine-tuning on MELD dataset.}
\label{tab:finetune-results}
\begin{tabular}{@{}cccc@{}}
\toprule
Fine-tuning Epoch & UA(\%) ↑ & WA(\%) ↑ & F1(\%) ↑ \\ \midrule
C$^2$SER (0 epoch) & 38.66    & 53.10    & 33.10    \\
3                  & 43.50    & 58.90    & 38.70    \\
6                  & 49.30    & 64.86    & 44.10    \\ \bottomrule
\end{tabular}
\end{table}

\subsection{Category Accuracy of Emotion2Vec-S and C$^2$SER}
\label{append:emo2vec_details}

\begin{figure}[t]
  \centering
  \includegraphics[width=\linewidth]{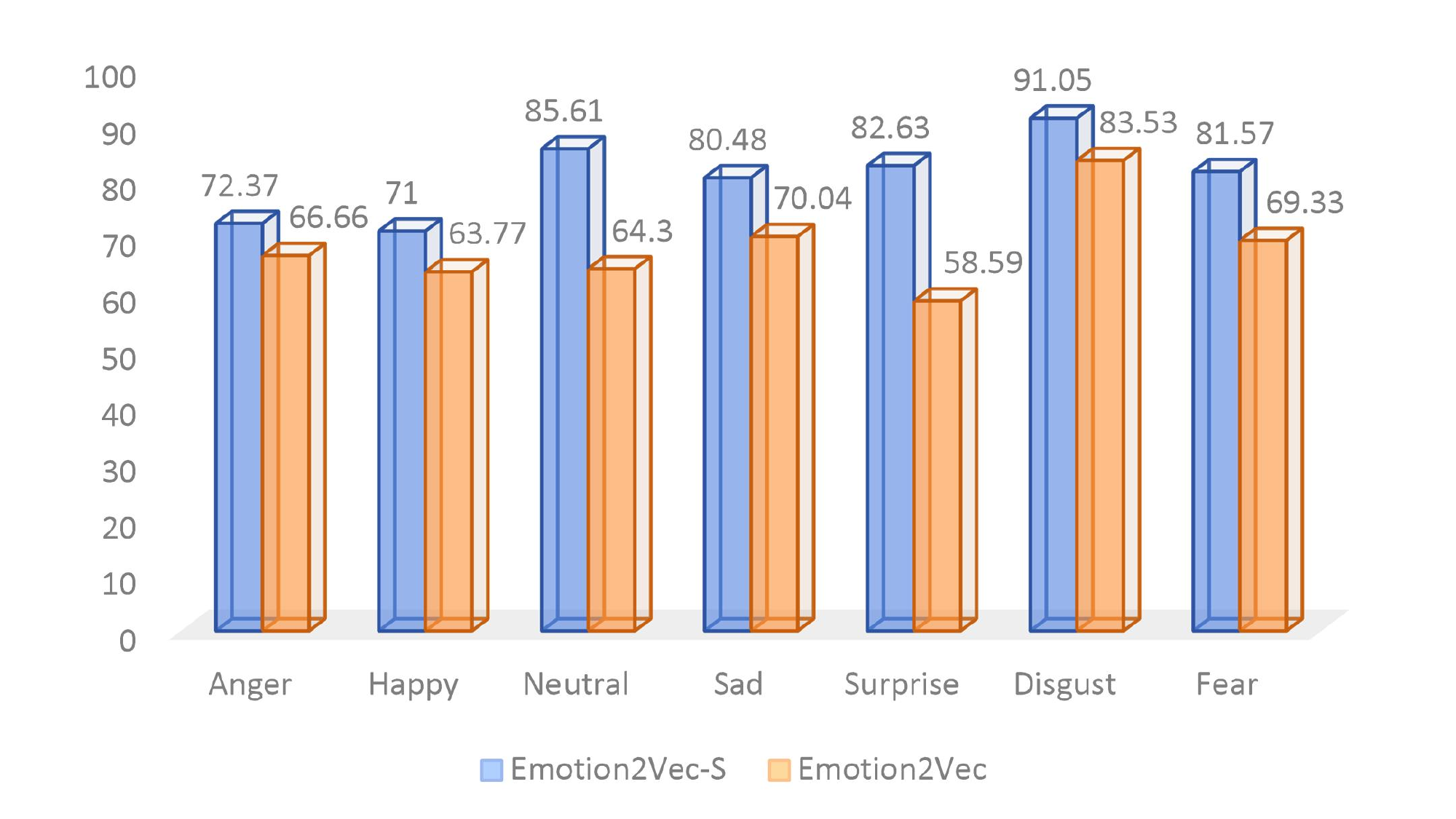}
  \caption {Category Accuracy (\%) of Emotion2Vec-S on the Emo-Emilia test set.}
  \label{fig-5}
\end{figure}

\begin{figure}[t]
  \centering
  \includegraphics[width=\linewidth]{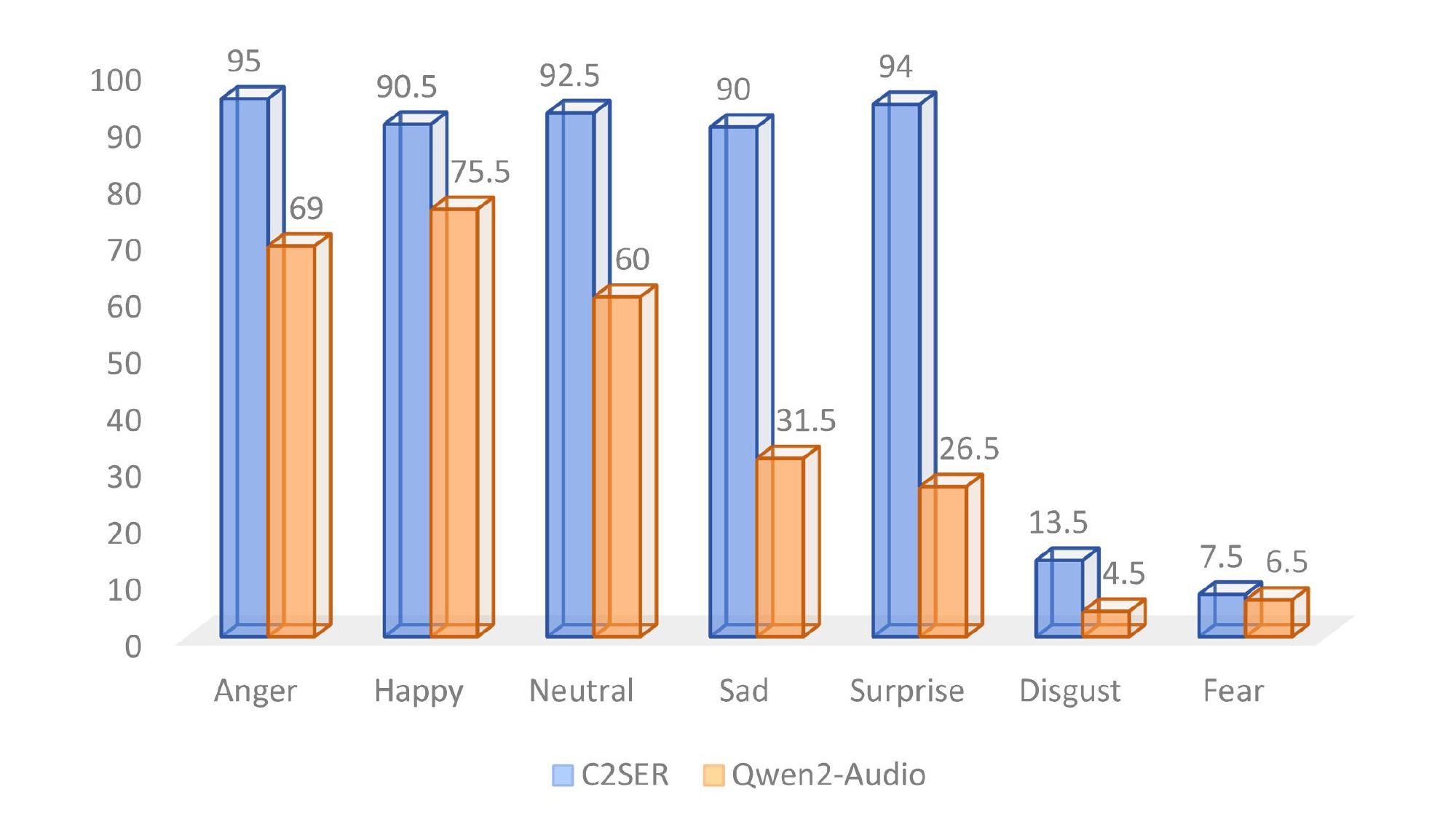}
  \caption {Category Accuracy (\%) of C$^2$SER on the Emo-Emilia test set.}
  \label{fig-6}
\end{figure}

We evaluate the category accuracy of Emotion2Vec-S using five-fold cross-validation, with 20\% of the training set used as the validation set. The average results across each fold are shown in Figure~\ref{fig-5}. Emotion2Vec-S outperforms Emotion2Vec in recognition accuracy for all emotion categories. Disgust achieves the highest recognition accuracy, while happiness has relatively lower recognition accuracy. In all cases, the accuracy of Emotion2Vec-S is higher than that of Emotion2Vec. Overall, the performance is relatively balanced across the emotions.

We evaluate the category accuracy of C$^2$SER through direct inference on the Emo-Emilia test set. The results are displayed in Figure~\ref{fig-6}. C$^2$SER achieves higher accuracy than Qwen2-audio across all emotion categories. The recognition accuracy for anger, happiness, neutral, sadness, and surprise is above 90\%, while the recognition accuracy for disgust and fear is below 20\%. This imbalance in performance is likely attributed to the skewed distribution of the training corpus, as shown in Figure~\ref{fig-3}.

\subsection{Effectiveness of CoT training}

To validate the effectiveness of the CoT training of C$^2$SER, we use the exact text instructions on Qwen2-Audio to conduct CoT inference. The comparison results are shown in Table~\ref{tab:cot-training}. Obviously, Qwen2-Audio is not well capable of CoT reasoning, whose performance is significantly lower than that of C$^2$SER. This result reveals that our CoT training boosts the reasoning capabilities of ALMs, enabling them to incorporate speech context for more accurate emotion recognition. Furthermore, Qwen2-Audio with implicit CoT performs worse than with explicit CoT, as it suffers from severe hallucinations that generate irrelevant results. This suggests that without explicit CoT training, implicit CoT fails to effectively guide reasoning and emotion recognition.

Having established the effectiveness of our training, we now analyze the specific advantages of our final Implicit CoT model. While its performance improvement over the Explicit CoT model is modest on the high-quality EMO-EMILIA test set (as seen in Table~\ref{tab:cot-training}), its primary value is demonstrated in its enhanced robustness, generalization, and efficiency. The improved robustness is empirically validated by the results in Table~\ref{tab:llm_result}, where Implicit CoT significantly outperforms Explicit CoT on more challenging and diverse datasets like M3ED and CASIA. This is because compressing the multi-step reasoning into a single, holistic step mitigates the risk of error propagation. Finally, the efficiency gain is inherent to the autoregressive process: Explicit CoT requires generating a long rationale (often $>40$ tokens), whereas Implicit CoT produces a concise expression ($<10$ tokens), guaranteeing an order-of-magnitude reduction in latency and making it viable for practical deployment.




\subsection{Ablation Study}

We conduct an ablation study to evaluate the contribution of each component in C$^2$SER. The experimental results are presented in Table~\ref{tab:ablation}.
Firstly, removing the Whisper encoder leads to a significant degradation in performance, with the model failing to converge during explicit CoT training due to the lack of semantic perception. 
Secondly, the model incorporating Emotion2Vec-S outperforms the version without it, demonstrating that acoustic perception is crucial for capturing emotional expressions effectively.
Finally, excluding CoT causes a substantial drop in performance. This result suggests the reasoning capability of C$^2$SER is improved after CoT training, which leads to a more accurate and stable emotion recognition.

In addition to evaluating each component, we further examine how C$^2$SER performs when fine-tuned on a specific dataset (MELD). As shown in Table~\ref{tab:finetune-results}, even a few epochs of fine-tuning can significantly boost the model’s performance on the target domain. However, since our primary objective is to ensure robust generalization across diverse emotional speech scenarios, we do not apply such dataset-specific fine-tuning in our main experiments.

\section{Conclusion and Future Work}

This paper proposes C$^2$SER, a novel audio-language model designed to address hallucinations in speech emotion recognition. Specifically, C$^2$SER introduces a contextual perception module of Whisper and Emotion2Vec-S, providing detailed semantic and acoustic information for the LLM decoder. Additionally, C$^2$SER introduces a chain-of-thought to incorporate speech context for emotion recognition, incentivizing reasoning capability. Furthermore, C$^2$SER proposes self-distillation, maintaining reasoning capability while minimizing error accumulation. Extensive experiments demonstrate that Emotion2Vec-S effectively captures emotion-related information, and C$^2$SER achieves an accurate and stable SER compared to existing models. 

Despite these advances, several key challenges open avenues for future research. A primary direction is to further enhance the model's generalization and real-world robustness. This involves not only curating more balanced and diverse training corpora to address performance variability, but also enriching the model's contextual understanding through multimodal inputs (e.g., visual cues) and systematically evaluating its sensitivity to interactive factors like prompt phrasing. Another crucial challenge is bridging the gap between advanced reasoning capabilities and deployment efficiency. While our 7B-parameter model with implicit CoT demonstrates strong performance, future work should explore model compression and specialized fine-tuning strategies. The goal is to create more efficient variants that strike a better balance between task-specific expertise and the model's inherent general-purpose language understanding, ultimately advancing the practical application of robust SER systems.

\bibliographystyle{IEEEtran}  
\bibliography{myreference}

\end{document}